\documentclass[10pt,english]{article}
\usepackage{times}
\usepackage[T1]{fontenc}
\usepackage[latin1]{inputenc}
\usepackage{amsmath}
\usepackage{amssymb}

\makeatletter
\usepackage{slashed}

\usepackage{babel}
\makeatother
\begin{document}

\title{\textbf{NEUTRINO MASSES WITH A SUITABLE PARAMETRIZATION IN THE PPF
3-3-1 GAUGE MODEL }}

\author{ADRIAN PALCU}

\date{\emph{Faculty of Exact Sciences - ''Aurel Vlaicu'' University Arad,
ComplexM, Elena Dragoi Str. 2, 310330 - Arad, Romania}}

\maketitle
\begin{abstract}
Plausible phenomenological consequences of the well-known Pisano-Pleitez-Frampton
3-3-1 model - such as the neutrino masses - are analyzed within the
solution provided by the exact algebraical approach - proposed several
years ago by Cot\u{a}escu - for gauge models with high symmetries.
We prove that a suitable parametrisation in the Higgs sector and a
redefinition of the three scalar triplets involved therein can lead
to realistic predictions for the lepton mass spectrum, while a minimal
number of coupling parameters are employed in the Yukawa sector. 

PACS numbers: 12.10.Dm; 12.60.Fr; 14.60.Pq, 12.60.Cn.

Key words: 3-3-1 gauge models, extended electroweak Higgs sector,
neutrino masses and mixings
\end{abstract}

\section{Introduction}

One of the most investigated extentions of the Standard Model (SM)
in the last decade is the so called Pisano-Pleitez-Frampton (PPF)
model \cite{key-1,key-2} based on the gauge group $SU(3)_{c}\otimes SU(3)_{L}\otimes U(1)_{Y}$
(in short ''3-3-1'') that - in its electroweak sector - undergoes
a spontaneous symmetry breakdown (SSB) up to the $U(1)_{em}$ electromagnetic
one. In the literature, the scalar sector of such a model assumes
an extended Higgs mechanism in two steps ($331\rightarrow321\rightarrow31$).
Three scalar triplets plus a scalar sextet \cite{key-3} are employed
in order to generate plausible masses for all the fermions and bosons
in the model. A detailed analysis of the most general Higgs potential
in 3-3-1 models that contains three scalar triplets and a scalar sextet
is worked out in Ref. \cite{key-4}. Note that PPF model is a particular
version of the rich class of 3-3-1 models, namely the one that predicts
exotic electric charges both in the boson sector ($\pm2e$) and the
quark sector ($\pm4e/3;\pm5e/3$). The fundamental fermion triplet
is the leptonic one and its three positions are occupied as follows:
left-handed charged lepton $l_{L}$, right-handed charged lepton $l_{R}$,
and corresponding left-handed neutrino $\nu_{lL}$. This structure
is identically triplicated for the well-kown generations: $e$ family,
$\mu$ family, and $\tau$ family respectively.

An exact algebraical approach for gauge models with high symmetries
$SU(n)_{L}\otimes U(1)_{Y}$ - subject to SSB - was proposed several
years ago by Cot\u{a}escu \cite{key-5}. The method displays a minimal
Higgs mechanism (mHm) in one step ($331\rightarrow31$) that finally
gives rise to the mass genrating Yukawa terms in unitary gauge and
allows for only one surviving neutral scalar - namely, the physical
Higgs field - just like in the SM. The results esentially depend on
a parametrisation in the Higgs sector of the model where a vector
space structure is imposed and the required number of Higgs multiplets
(vectors in different directions) is equal to the dimention of the
fundamental irreducible representation (irrep) of the electroweak
group. Any other necessary scalar representations (including the required
sextet) are obtained out of these multiplets by constructing certain
tensor-like products among them. At the same time, the correction
due to the mixing angle between the neutral gauge bosons is not needed
any more at the end of the calculus, since this task is performed
as a step of the method itself by means of a special generalized Weinberg
transformation (gWt). 

When applied to the PPF 3-3-1 model, the method supplies exact results
regarding the boson and lepton mass spectrum, the charged and neutral
currents in the model, as well as the possible neutrino mass patterns,
all being presented in Ref. \cite{key-6}. We strictly followed therein
the prescriptions of the general procedure, so that the mHm exhibits
one vev only. Therefore, the Yukawa sector calls for a plethora of
free coupling parameters. Here we develop those results in a more
suitable direction, by taking into account a redefinition of the scalar
fields that leads to a proper involvment of the scalar sector's parameters
in the vev splitting and thus to a decrement of the number of the
free coupling parameters in the Yukawa sector. All the previously
obtained results regarding the boson sector and the charges in the
model are not affected.

The paper is organized as follows. A brief review of the model is
presented in Sec.2 while Sec.3 contains our proposal for the possible
Yukawa terms to generate both Dirac and Majorana masses, dealing with
only three coupling parameters (a distinct one for each lepton family).
Sec.4 is devoted to our conclusions and some phenomenological predictions.

\section{Brief review of the model}

The Lagrangian denisity (Ld) of any gauge model that undergoes a SSB
must consist of several distinct terms, each describing one of the
following sectors: (i) the fermion sector, (ii) the gauge boson sector
and (iii) the Higgs sector, respectively. In adition, (iv) the Yukawa
sector must be employed in order to generate fermion masses, once
the SSB took place. For our purpose here, the latter one is of great
interest. Therefore, we begin by briefly presenting its ingredients
- namely, the fermion families and the scalar triplets - as irreps
of the 3-3-1 gauge group. Consequently, we foccus on the possible
ways of constructing the mass generating terms. In this respect, a
sextet is constructed in the scalar sector. The gauge sector consists
of two neutral bosons ($Z$, $Z^{\prime}$), the photon ($\gamma$),
two singly charged bosons ($W^{\pm}$,$V^{\pm}$) and a doubly-charged
one ($W^{\pm\pm}$). However, all the details regarding the gauge
sector - such as the boson mass spectrum and the charges of the particles
with respect to these bosons- are definitely established in Ref.\cite{key-6}.
The PPF 3-3-1 model displays the following anomaly-free particle content:

\textbf{Lepton families}\begin{equation}
\begin{array}{ccccc}
f_{\alpha L}=\left(\begin{array}{c}
e_{\alpha}^{c}\\
e_{\alpha}\\
\nu_{\alpha}\end{array}\right)_{L}\sim(\mathbf{1,3},0) &  &  &  & \left(e_{\alpha L}\right)^{c}\sim(\mathbf{1},\mathbf{1},-1)\end{array}\label{Eq.1}\end{equation}

\textbf{Quark families}

\begin{equation}
\begin{array}{ccc}
Q_{iL}=\left(\begin{array}{c}
J_{i}\\
u_{i}\\
d_{i}\end{array}\right)_{L}\sim(\mathbf{3,3^{*}},-1/3) &  & Q_{3L}=\left(\begin{array}{c}
J_{3}\\
-b\\
t\end{array}\right)_{L}\sim(\mathbf{3},\mathbf{3},+2/3)\end{array}\label{Eq.2}\end{equation}
\begin{equation}
\begin{array}{ccc}
(b_{L})^{c},(d_{iL})^{c}\sim(\mathbf{3},\mathbf{1},-1/3) &  & (t_{L})^{c},(u_{iL})^{c}\sim(\mathbf{3},\mathbf{1},+2/3)\end{array}\label{Eq.3}\end{equation}
\begin{equation}
\begin{array}{ccccccccc}
(J_{3L})^{c}\sim(\mathbf{3,1},+5/3) &  &  &  &  &  &  &  & (J_{iL})^{c}\sim(\mathbf{3,1},-4/3)\end{array}\label{Eq.4}\end{equation}
 with $\alpha=1,2,3$ and $i=1,2$. In the gauge group's iirreps displayed
above we assume, like in majority of the papers in the literature,
that the third generation of quarks transforms differently from the
other two ones. This could explain the unusual heavy masses of the
third generation of quarks, and especially the uncommon properties
of the top quark. The capital letters $J$ denote the exotic quarks
included in each family. The subscript $L$ means left-handed component,
while the numbers in parantheses are the representations and characters
with respect to the gauge group of the model. These fermion families
must be coupled through the scalar irreps in order to form mass generating
terms.

\textbf{Higgs triplets }

The Higgs sector consists of three scalar triplets which obey the
following irreps $\phi^{(1)}\sim(\mathbf{1},{\textbf{3}},-1)$, $\phi^{(2)}\sim(\mathbf{1},{\textbf{3}},1)$
and $\phi^{(3)}\sim(\mathbf{1},{\textbf{3}},0)$. The mHm prescribed
by the general method \cite{key-5} and formerly applied to the PPF
model \cite{key-6} involves the parameters 

\begin{equation}
\eta^{2}=(1-\eta_{0}^{2}){\textrm{diag}}\left(1-a,\frac{a(1-3\tan^{2}\theta_{W})}{2},\frac{a(1+3\tan^{2}\theta_{W})}{2}\right)\label{Eq.5}\end{equation}
 only in the boson mass aspectrum, since the single remaining vev
of the model $\left\langle \phi\right\rangle $ requires a lot of
new parameters $A$, $B$, $C$, $A^{\prime}$, $B^{\prime}$, $C^{\prime}$
- Yukawa coupling coefficients in the lepton Yukawa sector - in order
to generate adecquate masses. The aim of this paper is proving that
a more suitable approach can reduce at least three of the above Yukawa
couplings. 

In this respect one redefines the scalar triplets in the folloowing
manner:

\begin{equation}
\phi^{(i)}\rightarrow\eta^{(i)}\phi^{(i)}\label{Eq.6}\end{equation}
keeping at the same time the orthogonality condition (Eq.(27) in Ref.\cite{key-5})
for the new scalar triplets in order to avoid the Goldstone bosons
successively the SSB.

It is easy to check that this redefinition does not alter anyhow the
minimum condition (Eq.(33) in Ref.\cite{key-5}) for the scalar potential
(Eq.(32) in Ref. \cite{key-5}), since $\lambda$'s are still arbitrary
and they will account only for the Higgs mass (but this is not our
task here). 

The advantage of the present method is that of splitting the vev $\left\langle \phi\right\rangle $into
three vevs $\left\langle \phi^{(i)}\right\rangle =\eta^{(i)}\left\langle \phi\right\rangle $,
thus getting closer to the traditonal approaches to 3-3-1 models.
The trace condition in the parameter matrix Eq.(29) in Ref.\cite{key-5}
ensures that $\left\langle \phi\right\rangle ^{2}=\left\langle \phi^{(1)}\right\rangle ^{2}+\left\langle \phi^{(2)}\right\rangle ^{2}+\left\langle \phi^{(3)}\right\rangle ^{2}$

\section{Fermion masses}

\subsection{Charged lepton masses}

In the PPF 3-3-1 model, the charged leptons aquire their masses by
means of a scalar sextet \cite{key-3}, which is a compulsory ingredient
in the Yukawa Ld. We build here this scalar sextet out of the scalar
triplets already existing in the Higgs sector of the model as a tensor-like
product in the following manner: \begin{equation}
S=\phi^{-1}\left(\phi^{(1)}\otimes\phi^{(2)}+\phi^{(2)}\otimes\phi^{(1)}\right)\label{Eq.7}\end{equation}
 It plays the same role as the tensor blocks $\chi^{\rho\rho^{\prime}}$
in Eq.(16) in Ref.\cite{key-5}. Evidently, $S\sim(\mathbf{1},\mathbf{6},0)$
and thus the generating mass term in the charged leptons sector reads
\begin{equation}
G_{\alpha}{}\bar{f}_{\alpha L}Sf_{\alpha L}^{c}+H.c.\label{Eq.8}\end{equation}
 Hence, consequently the SBB, only positions (12) and (21) in Eq.(8))
will remain non-zero. For the redefined scalar fields, that is \begin{equation}
\left\langle S\right\rangle =\left(\begin{array}{ccc}
0 & \eta^{(1)}\eta^{(2)} & 0\\
\eta^{(1)}\eta^{(2)} & 0 & 0\\
0 & 0 & 0\end{array}\right)\langle\phi\rangle\label{Eq.9}\end{equation}
 All the charged fermions aquire thir masses through the above presented
coupling terms - Eq.(8), since all couplings due to $S$ get in the
unitary gauge the traditional Yukawa form: $G_{\alpha}\langle\phi\rangle\bar{e}_{\alpha L}e_{\alpha L}^{c}$(according
to a Dirac Lagrangian density put in the pure left form - see Appendix
B in Ref. \cite{key-5}). Therefore, one can identify the mass of
the charged lepton as \begin{equation}
m(e_{\alpha})=G_{\alpha}\eta^{(1)}\eta^{(2)}\left\langle \phi\right\rangle \label{Eq.10}\end{equation}
 Note that Eqs.(10) introduce $3$ more parameters $G_{\alpha}$ in
the model. Let them be: $A$ for $e$, $B$ for $\mu$ and $C$ for
$\tau$. 

In the literature on the 3-3-1 models, the Higgs triplets irreps are
commonly denoted by $\rho\sim\left(\mathbf{1,3},1\right)$, $\eta\sim\left(\mathbf{1,3},0\right)$
and $\chi\sim\left(\mathbf{1,3},-1\right)$ One has now to perform
a bijective mapping $(\left\langle \chi\right\rangle ,\left\langle \rho\right\rangle ,\left\langle \eta\right\rangle )\rightarrow(\left\langle \phi^{(1)}\right\rangle ,\left\langle \phi^{(2)}\right\rangle ,\left\langle \phi^{(3)}\right\rangle )$
in order to identify the scalar triplets from our method. Consequently,
one actualy deals with 3 possible distinct cases after SSB, since
Eq.(37) in Ref.\cite{key-5} allows us to perform a boost $U_{j\cdot}^{\cdot i}$
toward a new gauge that is equivalent to the unitary one. Therefore,
a simple permutation could well be taken into consideration (the boost
itself can perform it!), since in the unitary gauge Eq.(36) in Ref.\cite{key-5}
one could take $\delta_{i,j-1}$ or $\delta_{i,j+1}$ instead of $\delta_{i,j}$. 

At this point, our method offers three distinct cases for the possible
masses in the charged lepton sector, while keeping unchanged the order
in the parameter matrix $\eta$. The charged lepton mass can have,
with respect to the parameter order, the following possible values: 

\begin{itemize}
\item $m(e_{\alpha})=\frac{1}{2\sqrt{2}}G_{\alpha}\sqrt{(1-a)a(1-3\tan^{2}\theta_{W})}\left\langle \phi\right\rangle $
(case I) 
\item $m(e_{\alpha})=\frac{1}{2\sqrt{2}}G_{\alpha}\sqrt{(1-a)a(1+3\tan^{2}\theta_{W})}\left\langle \phi\right\rangle $
(case II) 
\item $m(e_{\alpha})=\frac{1}{4}G_{\alpha}a\sqrt{(1-9\tan^{4}\theta_{W})}\left\langle \phi\right\rangle $
(case III). 
\end{itemize}
Up to this stage each of these cases has two subcases for the choice
of the remaining two scalar triplets. As further steps, one has to
investigate the phenomenological aspects of each choice and rule out
the unsuitable ones.

It is worth to note that this kind of approach led to plausible phenomenological
predictions \cite{key-7,key-8} in the case of the 3-3-1 models with
right-handed neutrinos.

\subsection{Neutrino Mass Matrix }

Since the neutrino oscillations are an undisputable evidence \cite{key-9}
- \cite{key-15}, all the extentions of the SM must incorporate realistic
theoretical mechanisms for generating tiny masses in the neutrino
sector. There are two main lines in the literature to obtain these
tiny masses: (a) \textbf{see-saw mechanism} \cite{key-16} - \cite{key-18}
(see, for instance, Refs. \cite{key-19} - \cite{key-24} for its
particular realisations in 3-3-1 models) and (b) \textbf{radiative
corrections} (widely exploited in various variants of 3-3-1 models
\cite{key-25} - \cite{key-31}). 

We propose here neutrino mass terms at tree level in the Yukawa Ld
of the PPF model. The order of magnitude for these masses will be
a matter of tuning the free parameter $a$ of the model. We examine
in the following two distinct possibilities with different phenomenological
implications. The PPF 3-3-1 model allows for either Dirac or Majorana
masses in the neutrino sector, and even for both at the same time
(since the see-saw mechanism can occur). 

In order to minimize the number of free parameters in the Yukawa sector
one can assign a unique coupling to each lepton family.

\paragraph{Dirac neutrinos}

Assuming the existence of the right-handed neutrinos one can add in
the leptonic Yukawa sector of the PPF model presented above a supplementary
term of the form

\begin{equation}
G_{\alpha\beta}\bar{f}_{\alpha L}\eta^{*}\nu_{\beta R}+H.c\label{Eq.11}\end{equation}

After SSB such a canonical Yukawa term generates a pure Dirac neutrino
mass matrix:

\begin{equation}
M(\nu)=\frac{1}{2}\left(\begin{array}{ccc}
A & D & L\\
E & B & F\\
K & G & C\end{array}\right)\eta^{(\eta)}\left\langle \phi\right\rangle \label{Eq.12}\end{equation}
Obviously, the coupling constants are in our notation: $A=G_{ee}$,
$B=G_{\mu\mu}$, $C=G_{\tau\tau}$, $D=G_{e\mu}$, $E=G_{\mu e}$,
$F=G_{\mu\tau}$ , $G=G_{\tau\mu}$, $L=G_{e\tau}$, $K=G_{\tau e}$. 

At this point the three possible cases (determined by the vev alignment)
are expressed as a function depending on the sole parameter $a$. 

\begin{itemize}
\item $m(\nu_{\alpha})=\frac{1}{\sqrt{(1-a)}}\sqrt{\frac{(1+3\tan^{2}\theta_{W})}{(1-3\tan^{2}\theta_{W})}}m(e_{\alpha})$
(case I) 
\item $m(\nu_{\alpha})=\frac{1}{\sqrt{(1-a)}}\sqrt{\frac{(1-3\tan^{2}\theta_{W})}{(1+3\tan^{2}\theta_{W})}}m(e_{\alpha})$
(case II) 
\item $m(\nu_{\alpha})=2\frac{\sqrt{1-a}}{a}\frac{1}{\sqrt{(1-9\tan^{4}\theta_{W})}}m(e_{\alpha})$
(case III). 
\end{itemize}

\paragraph{Majorana neutrinos}

A second possibility is to introduce pure Majorana terms for neutrinos.
Under these circumstances, the leptonic Yukawa sector is completed
by a special term:

\begin{equation}
G_{\alpha\beta}{}\bar{f}_{\alpha L}\left[\phi^{-1}\left(\phi^{(\eta)}\otimes\phi^{(\eta)}\right)\right]Sf_{\beta L}^{c}+H.c.\label{Eq.13}\end{equation}
which develops the well-kown Yukawa shape in unitary gauge, succesively
the SSB. We notice that for the Majorana case, matrix $M$ is a symmetric
real one, with $D=E$, $F=G$, $L=K$. Hence, the Majorana neutrino
mass matrix reads:

\begin{equation}
M(\nu)=\frac{1}{4}\left(\begin{array}{ccc}
A & D & E\\
D & B & F\\
E & F & C\end{array}\right)\left(\eta^{(\eta)}\right)^{2}\left\langle \phi\right\rangle \label{Eq.14}\end{equation}

The three possible cases (depending on the vev alignment) exhibit
the following mass spectrum for Majorana neutrinos. 

\begin{itemize}
\item $m(\nu_{\alpha})=\frac{1}{2}\sqrt{\frac{a}{(1-a)}}\frac{(1+3\tan^{2}\theta_{W})}{\sqrt{(1-3\tan^{2}\theta_{W})}}m(e_{\alpha})$
(case I) 
\item $m(\nu_{\alpha})=\frac{1}{2}\sqrt{\frac{a}{(1-a)}}\frac{(1-3\tan^{2}\theta_{W})}{\sqrt{(1+3\tan^{2}\theta_{W})}}m(e_{\alpha})$
(case II) 
\item $m(\nu_{\alpha})=\frac{1-a}{a}\frac{1}{\sqrt{(1-9\tan^{4}\theta_{W})}}m(e_{\alpha})$
(case III). 
\end{itemize}

\paragraph{Phenomenological restrictions on parameter \textmd{$a$}}

Now, either for Dirac or for Majorana species of neutrinos, one can
examine each of the obained expressions and compare them to the available
experimental data \cite{key-32} regarding the order of magnitude
in the neutrino mass spectrum and to other particular features the
neutrino phenomenology exhibits \cite{key-33} - \cite{key-38}. Of
course, some of the expressions displayed above for the neutrino masses
will be ruled out by certain restrictive conditions imposed by phenomenological
reasons. We foccus in the following on a single criterion, namely
the order of magnitude for the neutrino masses. 

A great deal of experimental data confirm that phenomenological values
\cite{key-37,key-38} of neutrino masses $m(\nu_{\alpha})$ are severely
limited to a few eVs. Therefore, one remains finaly with only a few
acceptable cases out of all possible ones that our method on theoretical
grounds allows. Let us compute the sum of the neutrino masses. It
is nothing but the trace of the neutrino mass matrix.

\begin{equation}
\sum_{\alpha}m(\nu_{\alpha})=TrM(\nu)=m(\tau)\left[1+\frac{m(\mu)}{m(\tau)}+\frac{m(e)}{m(\tau)}\right]f(a,\theta_{W})\label{Eq.15}\end{equation}

Firts of all, we observe that one can neglect the small ratios $m(\mu)/m(\tau)\sim0.05$
and $m(e)/m(\tau)\sim0.0002$ in Eq. (15). Hence, the required sum
will be well approximated by:

\begin{equation}
\sum_{\alpha}m(\nu_{\alpha})\simeq m(\tau)f(a,\theta_{W})\label{eq.61}\end{equation}

Note that Dirac neutrinos exclude the cases (I) and (II), since these
cases supplies unacceptable values for the neutrino mass spectrum
which now is lower bounded by $m(e_{\alpha})\sqrt{(1-3\tan^{2}\theta_{W})/(1+3\tan^{2}\theta_{W})}$
which is far above the eVs domain. Therefore, in the case of pure
Dirac neutrinos only the case (III) has to be further investigated.
It favours values in the vicinity of $1$ for parameter $a$. For
such a parameter $a$ also Majorana neutrinos accept only the case
(III). 

If one wants to keep the free parameter $a$ in the vicinity of $0$
one observes that pure Dirac neutrinos are not compatible with any
of the above cases, while Majorana neutrinos could be compatible with
cases (I) and (II). 

Therefore, at this point one can say that by just tuning the parameter
$a$, the neutrino masses - either they are pure Dirac or pure Majorana
fields - could come out at viable values. 

The see-saw mechanism can be - with this assignment - naturally implemented
in the model. However, this issue will be analyzed in a future work.

\section{Concluding Remarks}

In this paper we have proved that the well-kown PPF 3-3-1 model can
be investigated from an algebraical viewpoint by just tuning a single
free parameter $a$. All the phenomenological consequences of the
model occur due to this parameter. 

For instance, if we take $\sin^{2}\theta_{W}\simeq0.231$ and $m(\tau)=1.777$
GeV (Patricle Data Group \cite{key-32}), then plausible Dirac neutrino
masses - in the range $\sim1$eV - occur only if $a\simeq1$, more
precisely, a value $a\simeq(1-10^{-16})$ is neccesary. The case (III)
is the only one compatible with such a setting. Under these circumstances,
the ''old'' bosons remain at their SM mass values \emph{i.e.} $m(Z)\simeq91.1$GeV
and $m(W)\simeq80.4$GeV \cite{key-32} while the ''new'' ones now
become - according to Eqs. (26) in Ref. \cite{key-6} - lighter: $m(Z^{\prime})\simeq63.76$GeV,
$m(U)\simeq17.8$GeV and $m(V)\simeq78.1$GeV. Quite the same boson
mass spectrum is obtained if neutrinos aquire Majorana masses, also
in the case (III) presented in previous section. These possibilities
arise at not a very high breaking scale $\left\langle \phi\right\rangle \simeq500$GeV. 

For the Majorana neutrinos, case (II) - and even case (I) - also seem
to be viable if and only if $a\simeq10^{-20}$ or less. In these cases
the boson mass spectrum - the same formulas (26) in Ref. \cite{key-6}
- looks like: $m(Z^{\prime})\simeq0.03\times10^{16}$GeV and $m(U)\simeq m(V)\simeq0.0001\times10^{16}$GeV
and the vev $\left\langle \phi\right\rangle $ of the model lies in
the GUT energies region ($10^{16}$GeV). This seems to be the price
paid in order to have good phenomenological results for all the known
particles in this 3-3-1 model, using only one free parameter. 

Further experimental investigations at LHC will precisely determine
the value of the new bosons masses. Only then one can decide which
case of the three provided by our method will be favoured. However,
one can imagine a suitable see-saw mechanism (to be presented in a
future work) or radiative mechanims deisgned to supply good neutrino
masses without resorting to such a fine tuning.

Furthermore, the mass squared differences for the solar and atmospheric
neutrinos along with their mixing angles can be performed. This task
has already been accomplished within this very method by the author
\cite{key-39} in the case of 3-3-1 models with right-handed neutrinos
with good predictions even for the absolute minimal mass in the neutrino
sector. Those results could well be implemented \cite{key-40} with
some small adjustments even into PPF model. The specific difference
resides in the manner in which the free parameter $a$ is involved
in the final expressions of the absolute masses. The squared difference
ratio remains also independent of this parameter.

\end{document}